\documentclass[11pt]{article}
\usepackage{epsfig}
\newcommand{\sss}{\vspace{.2in}}

\newcommand{\be}{\begin{equation}}
\newcommand{\ee}{\end{equation}}
\newcommand{\bea}{\begin{eqnarray}}
\newcommand{\eea}{\end{eqnarray}}
\newcommand{\sn}{{\rm sn}}
\newcommand{\cn}{{\rm cn}}
\newcommand{\dn}{{\rm dn}}
\newcommand{\sech}{{\rm sech}}

\begin{document}
\vspace{.2in}
~\hfill{\footnotesize UICHEP-TH/01-7,~~December 31, 2001}
\vspace{.5in}
\begin{center}
{\LARGE \bf Cyclic Identities Involving Jacobi Elliptic Functions}
\end{center}
\vspace{.5in}
\begin{center}
{\Large{\mbox{AVINASH KHARE}\footnote{Permanent address: Institute of Physics, Sachivalaya Marg, Bhubaneswar 751005, Orissa, India}  and
   \mbox{UDAY SUKHATME} 
 }}\\
{\it \large Department of Physics, University of Illinois at Chicago\\ Chicago, Illinois 60607-7059} 
\\
\end{center}
\vspace{1.4in}
{\bf {Abstract.}}
We state and discuss numerous mathematical identities involving Jacobi elliptic functions  
$~\sn \,(x,m)$, $~\cn \,(x,m)$, $~\dn \,(x,m)$, where $m$ is the elliptic modulus parameter. In all identities, the arguments of the Jacobi functions are separated by either $2K(m)/p$ or $4K(m)/p$, where $p$ is an integer and $K(m)$ is the complete elliptic integral of the first kind.  Each $p$-point identity of rank $r$ involves a cyclic homogeneous polynomial of degree $r$ (in Jacobi elliptic functions with $p$ equally spaced arguments) related to other cyclic homogeneous  polynomials of degree $r-2$ or smaller. Identities corresponding to small values of $p,r$ are readily established algebraically using standard properties of Jacobi elliptic functions, whereas identities with higher values of $p,r$  are easily verified numerically using advanced mathematical software packages.\\ 

\vspace{.2in}


\newpage
\noindent{\bf {1.  Introduction}}

\sss
\noindent The Jacobi elliptic functions $\sn \,(x,m),~\cn \,(x,m)$ and $\dn \,(x,m)$ with real elliptic modulus
parameter $m$ $( 0\leq m\leq 1)$ have been extensively studied and used in mathematics, science and engineering \cite{abr,gra}. Recently, while studying \cite{ks,kha9} the properties of quantum mechanical periodic potentials \cite{ars,mw}, we have discovered numerous mathematical identities involving Jacobi elliptic functions. The purpose of this paper is to tabulate, derive and discuss these identities. 

To the best of our knowledge, our results are not discussed in the mathematics literature. However, we did find that geometrical constructions called the ``poristic polygons of Poncelet" give rise to a few of our very simplest identities like eqs. (\ref{E3}) and (\ref{E32}) involving just the Jacobi elliptic functions $\dn(x,m)$. For a discussion of this geometrical approach, see references \cite{gre, hal}. 

Our new identities play a crucial role in obtaining a large class of novel periodic solutions of the Korteweg-de Vries (KdV) and modified Korteweg-de Vries equations \cite{kha6}, the nonlinear Schr{\" o}dinger and KP equations, the sine-Gordon equation, as well as the $\lambda \phi^4$ model \cite{kha8}. 
The solutions obtained for the KdV equation \cite{kha6} all correspond to one gap periodic potentials. This process can be generalized to obtain new solvable periodic potentials with a finite number of band gaps \cite{kha9}.

If $K(m)$ denotes the complete elliptic integral of the first kind, the elliptic functions $\sn \,(x,m)$ and $\cn \,(x,m)$ have real periods $4K(m)$, whereas $\dn \,(x,m)$ has a period $2K(m)$. The $m =0$ limit gives $K(0)=\pi /2$ and trigonometric functions: $\sn(x,0)=\sin x, ~\cn(x,0)=\cos x, ~\dn(x,0)=1$. The $m \rightarrow 1$ limit gives $K(1) \rightarrow \infty$ and hyperbolic functions: $\sn(x,1) \rightarrow \tanh x, ~\cn(x,1) \rightarrow \sech \,x, ~\dn(x,1) \rightarrow \sech\, x$. Therefore, our new identities for Jacobi elliptic functions can be thought of as generalizations to arbitrary $m$ of identities involving trigonometric and hyperbolic functions.\sss  

\noindent{\bf {2.  Description of the identities}}

\sss
\noindent In all the identities discussed in this paper, the arguments of the Jacobi 
functions are separated by either $2K(m)/p$ or $4K(m)/p$, where $p$ is an 
integer ($p \ge 2$) depending on whether the left hand side of the identity is a periodic function of period $2K(m)/p$ or $4K(m)/p$. For any given choice of $p$, we define the quantities $s_i$,~$c_i$ and $d_i$ as follows:
\be\label{E1}
~ s_i \equiv \sn[x + \frac{2(i-1)K(m)}{p},m]~~,
~ c_i \equiv \cn[x + \frac{2(i-1)K(m)}{p},m]~~,
~ d_i \equiv \dn[x + \frac{2(i-1)K(m)}{p},m]~~.
\ee 
Similarly, we define
\be\label{E2}
{\tilde s}_i \equiv \sn[x+\frac{4(i-1)K(m)}{p},m]~;~{\tilde c}_i \equiv \cn[x+\frac{4(i-1)K(m)}{p},m]~,~
{\tilde d}_i \equiv \dn[x+\frac{4(i-1)K(m)}{p},m]~.
\ee

Each $p$-point identity which we discuss will involve a cyclic homogeneous polynomial of degree $r$ (in Jacobi elliptic functions with $p$ equally spaced arguments) expressed as a linear combination of other cyclic homogeneous polynomials of degree $r-2n$, where $1 \le n \le \frac{r}{2}$. We designate this to be a $p$-point identity of rank $r$.

Let us consider a few examples to clarify the terminology and establish the notation. A simple 4-point identity of rank 2 is
\be\label{E3}
d_1 d_2 + c.p. \equiv
d_1 d_2 + d_2 d_3 + d_3 d_4+ d_4 d_1= A~,
\ee
where we have used the notation ``$+~c.p.$" to denote cyclic permutations of the indices $1,2,\ldots,p$. Later, we have also used the notation ``$-~c.p.$" to denote cyclic permutations with alternating positive and negative signs. The quantities in eq. (\ref{E3}) are
\be\label{E4}
d_1 \equiv \dn(x,m)~,~d_2 \equiv \dn(x+K(m)/2,m)~,~d_3 \equiv \dn(x+K(m),m)~,~d_4 \equiv \dn(x+3K(m)/2,m)~.
\ee
Setting $x=0$, the constant $A$ can be computed to be $A = 2t (1+t^2)$
where
\be\label{E5}
t \equiv \dn(K(m)/2,m) =(1-m)^{1/4}~.
\ee
Similarly, two examples of 3-point identities of rank 2 and rank 3 are
\be\label{E6}
{\tilde c}_1 {\tilde c}_2 + c.p. = -\frac{q(q+2)}{(1+q)^2}~~,~~
{\tilde c}_1 {\tilde d}_2 {\tilde d}_3 + c.p. = -q^2({\tilde c}_1 + {\tilde c}_2 + {\tilde c}_3)~,
\ee
where
\be\label{E7}
q \equiv \dn(2K(m)/3,m)~,
\ee
and the arguments are $x, x+4K(m)/3$ and $x+8K(m)/3$ respectively.
Many more examples are given in Tables 1, 2 and 3. Identities of rank 2 are given in Table 1, identities of rank 3 are given in Table 2 and some examples of identities of rank 4 or greater are displayed in Table 3.

Although $x$-independent constants like $A$ do depend on the number of points $p$, the rank $r$, the modulus parameter $m$, and the specific identity involved, for simplicity, we do not usually exhibit these dependences explicitly. In fact, the symbols $A$, $B$ and $C$ appearing in the identities given in Tables 1, 2 and 3 are just meant to denote generic constants. They do not all have the same values.

For illustrative purposes, we now outline the proof of the $p$-point identity
\be\label{E8}
d_1 d_2 + c.p. =A~,
\ee 
which is true for any integer value $p > 1$. The left hand side of this identity contains $p$ terms. The proof for $p=2$ is trivial, since it is well-known that  
$d_1 d_2 \equiv \dn(x,m)~ \dn(x+K(m),m) = \sqrt{1-m}$  \cite{abr,gra}. For $p=3$, one needs to compute $d_1 d_2 +d_2 d_3 + d_3 d_1 \equiv \dn(x,m)~ \dn(x+2K(m)/3,m)+\dn(x+2K(m)/3,m)~ \dn(x+4K(m)/3,m)+\dn(x+4K(m)/3,m) ~\dn(x,m)$.
This can be accomplished by algebraic simplification after using the addition theorem \cite{abr,gra}
\be\label{E9}
\dn(u+v)=(\dn u ~\dn v -m\, \sn u ~\cn u ~ \sn v ~\cn v )/(1-m~\sn^2 u~ \sn^2 v)~~.
\ee
The result is the constant $A$ for the $p=3$ case. One gets $A = q(q+2)$, where $q$ has been defined in eq. (\ref{E7}).

Similarly, the result for $p=4$ has already been discussed following 
eq. (\ref{E3}). In principle, an analogous algebraic procedure can be used 
for any value of $p$, but the algebra becomes increasingly lengthier. We have therefore 
verified identity (\ref{E8}) numerically  using the advanced mathematical 
software package Maple. Note that for any chosen value of $p$, the constant 
$A$ equals $p$ in the limit $m=0$  and vanishes for $m \rightarrow 1$.

Given any $p$-point identity of rank $r$, one way of generating a new $p$-point identity of rank $r+1$ is by differentiation and use of the well-known formulas $$\frac{d}{dx} \sn(x,m) = \cn(x,m)~\dn(x,m)~~,~~\frac{d}{dx} \cn(x,m) = -\sn(x,m)~\dn(x,m)~~,$$
\be\label{E10}
\frac{d}{dx} \dn(x,m) = -m\,\sn(x,m)~\cn(x,m).
\ee 
For example, differentiation of the $p$-point rank 2 identity (\ref{E8}) yields the rank 3 identity
\be\label{E11}
s_1 c_1 (d_2 + d_p) + c.p. = 0~,
\ee
which reduces to the well-known trigonometric identity
\be\label{E12}
\sum_{i=1}^p \sin [2x + \frac {2(i-1)\pi}{p}]=0
\ee
in the limit $m=0$.

Another $p$-point rank $r$ identity of interest for $r \le p$ is
\be\label{E13}
d_1 d_2 \ldots d_r + c.p. = ~A ~~~~~~~~~~~ (r~ {\rm even})~,
\ee
\be\label{E14}
d_1 d_2 \ldots d_r + c.p. = B\sum_i^p d_i ~~~~(r~ {\rm odd})~.
\ee
One also has similar identities involving ${\tilde s}_i$ or ${\tilde c}_i$ (instead of $d_i$) for any odd value of $p$.  
All these identities have the remarkable property of reducing the degree of the polynomial in the Jacobi functions from $r$ to 0 (1) depending on whether $r$ is even (odd). For small values of $p$ and $r$, the constants $A,B$ in eqs. (\ref{E13}) and (\ref{E14}) are easily evaluated. Some results are $$A(p=2,r=2)=2\sqrt{1-m}~~,~~A(p=3,r=2) = q(q+2)~~,~~A(p=4,r=2) = 2t(1+t^2)~,$$$$B(p=3,r=3)=3(\frac{m}{1-q^2}-1)~~,~~  B(p=4,r=3)=\sqrt{1-m}~,$$
\be\label{E15}
A(p=4,r=4) = 4(1-m)~.
\ee
For the special limiting cases $m=0$ and $m=1$, one gets
\be\label{E16}
A(m=0,p,r) = p~,~B(m=0,p,r) = 1~,~ A(m=1,p,r) = B(m=1,p,r) = 0~.
\ee

Another way of obtaining additional identities is by manipulating established identities.
For example, for  $p=3$ and $r=3$, eq. (\ref{E14}) is
\be\label{E17}
d_1 d_2 d_3 = \frac{B}{3} (d_1 + d_2 + d_3)~,
\ee
where $B \equiv B(p=3,r=3)$ is as given by eq. (\ref{E15}).
Squaring identity (\ref{E8}) for $p=3$ and using eqs. (\ref{E15}) and 
(\ref{E17}) yields
the new identity
\be\label{E18}
d_1^2 d_2^2 +c.p. = -2(\frac{m}{1-q^2}-1)\sum_{i=1}^3 d_i^2 + [(1-q^2)^2+\frac{6m}{1-q^2}-3-4m)]~.
\ee
A similar identity is also true for any $p$, and in fact we have used it in a crucial manner for obtaining new periodic solutions of the KdV equation \cite{kha6}. However, to establish this $p$-point identity, one needs a generalization of identities (\ref{E8}) and (\ref{E14}). The generalized identities are
\be\label{E19}
d_1 d_n + c.p. =A~, ~~(n=2,3,4,...)~,
\ee
and 
\be\label{E20}
d_1 d_{j_1} d_{j_2} + c.p. = B \sum_{i=1}^p d_i , ~~(1 < j_1 < j_2 \le p)~, 
\ee  
which we have verified to be true both algebraically and numerically using Maple for many specific choices of the integers $n,j_1,j_2,p$.\sss

\noindent{\bf {3.  Discussion and comments}}

\sss
\noindent By the techniques described in the previous section, we have obtained a large number of new identities, many of which are displayed in Tables 1, 2 and 3. It should be noted that the modulus parameter $m$ is not transformed and remains unchanged in all identities. Although it is not easy to give a complete systematic classification, we can comment on some general properties.\sss

\noindent(i) For any identity of rank $r$, the left hand side is a cyclic homogeneous polynomial expression of degree $r$ with $p$ terms.\sss

\noindent(ii) If the polynomial on the left hand side is periodic with period $2K(m)/p~[4K(m)/p]$, then the identity involves arguments spaced by $2K(m)/p~[4K(m)/p]$.\sss 

\noindent(iii) The right hand side involves polynomials of rank 
$r-2$, $r-4$,... which are ``irreducible", some examples being 
$\sum d_i, \sum {\tilde s}_i, \sum {\tilde c}_i, \sum {\tilde c}_i 
{\tilde d}_i, \sum {\tilde s}_i {\tilde d}_i, \sum c_i s_i, \sum c_i s_i d_i$, 
etc. and all these irreducibles multiplied by $d_i^{2n}$ where $n = 1,2,...$.\sss
 
\noindent(iv) In general, many of the identities of higher rank can be obtained from those of lower rank by either differentiation or algebraic manipulation. 
Similarly, many of the identities of a given rank $r~(r > 2$) can be derived
from lower rank identities as well as a few identities of the same rank. For 
example, for $p=3$, using the identities of rank 2 and three of the rank 3 
identities as given by eqs. (\ref{E32}) to (\ref{E34}) 
one can obtain all other identities of 
rank 3 as given in Table 2.\sss 

\noindent(v) The generic constants $A,\, B,\, C$ in any identity can be 
determined by choosing specific, convenient values of $x$ in the arguments. 
The value $x=0$ is a good choice in many cases. Note that for $p \le 4$, 
we have given explicit values for all the constants appearing in the identities - for 3-point identities, all constants are expressed in terms of $q \equiv \dn(2K(m)/3,m)~$, and for 4-point identities, all constants are expressed in terms of $t \equiv \dn(K(m)/2,m) =(1-m)^{1/4}~$. In writing the constants, we have made frequent use of the relationship $q^4+2q^3+(m-1)(2q+1)=0$.\sss

\noindent(vi) Some identities for even values of $p$ involve alternating 
positive and negative signs. The symbol ``$-~c.p.$" in these identities 
refers to cyclic permutations with alternating signs. Many of these 
identities, like $d_1^2(d_2+d_p)-d_2^2(d_3+d_1)\cdots-d_{p}^2(d_1+d_{p-1})
=A(d_1-d_2+\cdots-d_p)$, play a crucial role in determining band edge wave 
functions of solvable quantum mechanical periodic potentials \cite{kha9}.
For $p=4$ this identity is easily derived by starting from identity 
(\ref{E3}) with $A = 2t(1+t^2)$ and $t=(1-m)^{1/4}$. On multiplying 
both sides of this identity by $(d_1 -d_2 +d_3 -d_4)$ and using the relations
$d_1 d_3 = d_2 d_4 = \sqrt{1-m}$, we immediately obtain the 4-point identity
\be\label{E21}
d_1^2 (d_2 +d_4) - c.p. = 2t(1+t+t^2) (d_1 - c.p.)~.
\ee

\noindent(vii) It should be noted that our identities involve cyclic permutations $\pm c.p.$ of terms which have no clockwise or anticlockwise ``handedness".
For example, for even $p$, there is no identity of the type
$(d^2_1 d_2 - c.p.)$ proportional to 
$(d_1 -c.p.)$, since the term $d^2_1 d_2$ has a clockwise handedness. It is only when one adds on an anticlockwise handed term $d^2_1 d_p$ that the
the combination [$d_1^2 (d_2 +d_p) - c.p.$] is indeed proportional to
($d_1-c.p.)$. \sss  
 
\noindent(viii) In the limit $m \rightarrow 0$, one recovers many known non-trivial trigonometric identities. In the limit $m \rightarrow 1$, since the period $K(1) \rightarrow \infty$, one usually gets trivial hyperbolic function identities. Both these limits serve as a useful check on all the new identities involving Jacobi elliptic functions obtained in this paper. Of course, as mentioned previously, software packages like Maple or Mathematica quickly provide confirmation of any identity to typically eight digit accuracy.\sss

\noindent(ix) Identities for a given value of $p$, contain identities of 
the factors of $p$ as special cases. For example, for even $p$, only half of 
$d_1,...,d_p$ are independent since they satisfy identities 
\be\label{E22}
d_1 d_{\frac{p+2}{2}} = ... = d_{\frac{p}{2}} d_p = \sqrt{1-m}~,
\ee
coming from $p=2$. Similarly, the full list of $p=6$ identities contains 
$p=2,3$ identities. For example, $d_1 d_4 = d_2 d_5 = d_3 d_6 
=\sqrt{1-m}$ and similarly $d_1 d_3 + d_3 d_5 +d_5 d_1 = d_2 d_4 + d_4 d_6
+d_6 d_2 = q^2 + 2q$, where $q$ is as given by eq. (\ref{E7}). \sss

\noindent(x) It should be noted that in many applications like finding new solutions of the KdV equation \cite{kha6}, the identities needed involve summations over all combinations of many (say two) indices $i,j=1,\ldots,p$. These combinations correspond to the sum of several cyclic identities discussed in the tables.\sss

\noindent(xi) In this paper, we have concentrated our attention on cyclic identities in which the arguments are separated by fractions of the periods $2K(m)$ or $4K(m)$ on the real axis. However, each one of our identities can be easily translated into a corresponding one in which the arguments are separated by fractions of the periods $i2K'$ or $i4K'$ on the imaginary axis, where $K' \equiv K(1-m)$. For example, the simple 2-point identity $\dn(x,m)\,\dn(x+K(m),m)= \sqrt{1-m}$ translates to the new identity $\sn(u,m)\,\sn(u+iK',m)= 1/\sqrt{m}$. The general procedure consists of first replacing $m$ by $1-m$ [which in alternative standard notation \cite{gra} corresponds to replacing $k \equiv \sqrt{m}$ by $k' \equiv \sqrt{1-m}$ and $K(m)$ by $K'$], then using the well-known results 
$$
\sn(x,1-m)=\frac{i\,\cn(ix+K,m)}{\sqrt{1-m}~\,\sn(ix+K,m)}~~,~~\cn(x,1-m)=\frac{\dn(ix+K,m)}{\sqrt{1-m}~\,\sn(ix+K,m)}~~,
$$
$$
\dn(x,1-m)=\frac{1}{\sn(ix+K,m)}~~,
$$
and finally changing to a new variable $u = ix+K$.
\sss

In conclusion, even though Jacobi elliptic functions have been studied for approximately two centuries, it is exciting to discover new cyclic identities connecting them. What makes our results doubly exciting is that the identities play a vital role in the study of periodic potentials \cite{kha9} and in yielding new solutions of nonlinear differential equations of physical interest \cite{kha6}. \sss

\noindent{\bf {Acknowledgment}}

\sss
\noindent One of us (A.K) thanks the Department of Physics at the University of Illinois at Chicago for hospitality. This work was supported in part by the U.S. Department of Energy under grant  FG02-84ER40173.

\newpage
\noindent{\bf Table 1: Identities of rank 2.} The symbols $A$ in eqs. (\ref{E27}) and (29) are used generically to denote constants independent of $x$; the constants  
are in general all different. 

\sss
\noindent{\bf p = 2:}
\be\label{E23}
d_1 d_2 = \sqrt{1-m}
\ee
\sss
\noindent{\bf p = 3:}  $~~~[q \equiv \dn(2K(m)/3,m)]$
\be\label{E24}
d_1 d_2 + c.p. = q(q+2)~~,~~{\tilde c}_1 {\tilde c}_2 + c.p. =  
\frac{-q(q+2)}{(1+q)^2}~~,~~{\tilde s}_1 {\tilde s}_2 + c.p. = \frac{1}{m} (q^2-1)
\ee
\be\label{E25}
{\tilde c}_1 ({\tilde d}_2+{\tilde d}_3)+c.p.= {\tilde s}_1 ({\tilde d}_2+{\tilde d}_3)+c.p.= {\tilde c}_1 ({\tilde s}_2+{\tilde s}_3)+c.p.= 0
\ee
\sss
\noindent{\bf p = 4:} $~~~[t \equiv \dn(K(m)/2,m) =(1-m)^{1/4}]$
\be\label{E26}
d_1 d_3 = d_2 d_4 = \sqrt{1-m}~~,~~
d_1 d_2 + c.p. = 2t (1+t^2)
\ee
\sss
\noindent{\bf p = Even Integer:}
\be\label{E27}
d_1 d_2 + c.p. = A~~,~~d_1 d_3 + c.p. = A~~,~\ldots~,~~d_1 d_{\frac{p}{2}} + c.p. = A 
\ee
\be\label{E28}
d_1 d_{\frac{p}{2} +1} = d_2 d_{\frac{p}{2} +2} = \cdots = d_{\frac{p}{2}} d_{p} = \sqrt{1-m}
\ee
\sss
\noindent{\bf p = Odd Integer:}
\bea
d_1 d_2 &+& c.p. = A~,~{\tilde c}_1 {\tilde c}_2 + c.p. = A~,~{\tilde s}_1 {\tilde s}_2 + c.p. = A\nonumber \\
&\vdots&\nonumber \\
d_1 d_{\frac{p+1}{2}} &+& c.p. = A~,~{\tilde c}_1 {\tilde c}_{\frac{p+1}{2}} + c.p. = A~,~{\tilde s}_1 {\tilde s}_{\frac{p+1}{2}} + c.p. = A \\
{\tilde c}_1 ({\tilde d}_2+{\tilde d}_p)&+& c.p. =0~,~{\tilde s}_1 ({\tilde d}_2 +{\tilde d}_p)+ c.p. =0~,~{\tilde c}_1 ({\tilde s}_2 +{\tilde s}_p)+ c.p. =0 \nonumber\\
&\vdots&\nonumber \\
{\tilde c}_1 ({\tilde d}_{\frac{p+1}{2}}+{\tilde d}_{\frac{p+3}{2}})&+& c.p. =0~,~{\tilde s}_1 ({\tilde d}_{\frac{p+1}{2}}+{\tilde d}_{\frac{p+3}{2}})+ c.p. =0~,~{\tilde c}_1 ({\tilde s}_{\frac{p+1}{2}}+{\tilde s}_{\frac{p+3}{2}})+ c.p. =0
\eea

\newpage
\noindent{\bf Table 2: Identities of rank 3.} The symbols $A$ in eqs. (\ref{E58}) through (\ref{E65}) are used generically to denote constants independent of $x$; the constants  
are in general all different. 

\sss
\noindent{\bf p = 2:}
\be\label{E31}
d_1^2 d_2 \pm d_2^2 d_1 = \sqrt{1-m}\,(d_1 \pm d_2)~~,~~c_1s_1d_2+c_2s_2d_1=0
\ee
\sss
\noindent{\bf p = 3:}  $~~~[q \equiv \dn(2K(m)/3,m)]$
\be\label{E32}
d_1d_2d_3=\frac{(q^2+m-1)}{1-q^2}(d_1+d_2+d_3)
\ee
\be\label{E33}
{\tilde c}_1{\tilde c}_2{\tilde c}_3=\frac{q^2}{1-q^2}({\tilde c}_1 +{\tilde c}_2 + {\tilde c}_3) 
\ee
\be\label{E34}
{\tilde s}_1{\tilde s}_2{\tilde s}_3= \frac{-1}{1-q^2} ({\tilde s}_1 
+ {\tilde s}_2 + {\tilde s}_3 )
\ee
\be\label{E35}
{\tilde c}_1({\tilde s}_2{\tilde d}_3+{\tilde s}_3{\tilde d}_2)+c.p. =0
\ee
\be\label{E36}
{\tilde c}_1{\tilde d}_2{\tilde d}_3+c.p.=-q^2({\tilde c}_1+{\tilde c}_2+{\tilde c}_3)
\ee
\be\label{E37}
m{\tilde c}_1{\tilde s}_2{\tilde s}_3+c.p.=
-(1+q)^2({\tilde c}_1+{\tilde c}_2+{\tilde c}_3)
\ee
\be\label{E38}
{\tilde s}_1{\tilde d}_2{\tilde d}_3+c.p.=
\frac{(2q^3+3q^2-2q+3m-3)}{1-q^2}({\tilde s}_1+{\tilde s}_2+{\tilde s}_3)
\ee
\be\label{E39}
{\tilde s}_1{\tilde c}_2{\tilde c}_3+c.p.=\frac{q(q+2)}{1-q^2}({\tilde s}_1+{\tilde s}_2+{\tilde s}_3)
\ee
\be\label{E40}
{\tilde d}_1{\tilde c}_2{\tilde c}_3+c.p.=
\frac{-q^2}{(1+q)^2}({\tilde d}_1+{\tilde d}_2+{\tilde d}_3)
\ee
\be\label{E41}
m{\tilde d}_1{\tilde s}_2{\tilde s}_3+c.p.=
\frac{(-q^3-q^2+q+1-2m)}{1+q}({\tilde d}_1+{\tilde d}_2+{\tilde d}_3)
\ee
\be\label{E42}
{\tilde d}_1({\tilde d}_2{\tilde c}_2+{\tilde d}_3{\tilde c}_3)+c.p.=
2q(q+1)({\tilde c}_1+{\tilde c}_2+{\tilde c}_3)
\ee
\be\label{E43}
m{\tilde s}_1({\tilde s}_2{\tilde c}_2+{\tilde s}_3{\tilde c}_3)+c.p.=
2q(q+1)({\tilde c}_1+{\tilde c}_2+{\tilde c}_3)
\ee
\be\label{E44}
{\tilde d}_1({\tilde d}_2{\tilde s}_2+{\tilde d}_3{\tilde s}_3)+c.p.=
\frac{2(q^2+2q-m+1)}{1+q}({\tilde s}_1+{\tilde s}_2+{\tilde s}_3)
\ee
\be\label{E45}
{\tilde c}_1({\tilde c}_2{\tilde s}_2+{\tilde c}_3{\tilde s}_3)+c.p.
=\frac{-2q(q+2)}{(1+q)(1-q^2)}({\tilde s}_1+{\tilde s}_2+{\tilde s}_3)
\ee
\be\label{E46}
{\tilde c}_1({\tilde c}_2{\tilde d}_2+{\tilde c}_3{\tilde d}_3)+c.p.=
\frac{-2q}{(1+q)^2}({\tilde d}_1+{\tilde d}_2+{\tilde d}_3)
\ee
\be\label{E47}
m{\tilde s}_1({\tilde s}_2{\tilde d}_2+{\tilde s}_3{\tilde d}_3)+c.p.=
\frac{2(q^3+q^2-q-1+m)}{1+q}({\tilde d}_1+{\tilde d}_2+{\tilde d}_3)
\ee
\be\label{E48}
d_1^2(d_2+d_3)+c.p. = \frac{2(q-m+1)}{1+q}(d_1+d_2+d_3)
\ee
\be\label{E49}
{\tilde c}_1^2({\tilde c}_2+{\tilde c}_3)+c.p. = 
\frac{-2(q-m+1)}{m}({\tilde c}_1+{\tilde c}_2+{\tilde c}_3)
\ee
\be\label{E50}
~~m{\tilde s}_1^2({\tilde s}_2+{\tilde s}_3)+c.p. = 
\frac{2(q^3+q^2-q+mq+2m-1)}{1-q^2}({\tilde s}_1+{\tilde s}_2+{\tilde s}_3)
\ee
\be\label{E51}
c_1s_1(d_2+d_3)+c.p. = 0~~,~{\tilde c}_1{\tilde d}_1({\tilde s}_2+{\tilde s}_3)+c.p. = 0~~,~{\tilde d}_1{\tilde s}_1({\tilde c}_2+{\tilde c}_3)+c.p. = 0
\ee
\sss
\noindent{\bf p = 4:} $~~~[t \equiv \dn(K(m)/2,m) =(1-m)^{1/4}]$
\be\label{E52}
d_1d_2d_3 \pm d_2d_3d_4+d_3d_4d_1\pm d_4d_1d_2 = \sqrt{1-m}\,(\pm d_1 +d_2 \pm d_3 + d_4)
\ee
\be\label{E53}
d_1^2(d_2+d_4) \pm d_2^2(d_3+d_1)+d_3^2(d_4+d_2)\pm d_4^2(d_1+d_3) = 2t(1 \mp t +t^2)(d_1 \pm d_2 +d_3 \pm d_4)
\ee
\be\label{E54}
d_1^2 d_3 \pm d_2^2 d_4 +d_3^2 d_1 \pm d_4^2 d_2 = \sqrt{1-m}\, (d_1 \pm d_2 + d_3 \pm d_4)
\ee
\be\label{E55}
c_1s_1(d_2+d_4)+c.p. = 0~~,~c_1s_1d_3+c_3s_3d_1 = 0~~,~c_2s_2d_4+c_4s_4d_2= 0
\ee
\sss
\noindent{\bf p = Even Integer:}
\be\label{E56}
c_1s_1(d_2+d_p)+c.p.=0~,~c_1s_1(d_3+d_{p-1})+c.p.=0~,\cdots,~c_1s_1(d_{\frac{p}{2}}+d_{\frac{p}{2}+2})+c.p.=0
\ee
\be\label{E57}
c_1s_1d_{\frac{p}{2}+1}+c_{\frac{p}{2}+1}s_{\frac{p}{2}+1}d_1=c_2s_2d_{\frac{p}{2}+2}+c_{\frac{p}{2}+2}s_{\frac{p}{2}+2}d_2=\cdots=c_{\frac{p}{2}}s_{\frac{p}{2}}d_{p}+c_{p}s_{p}d_{\frac{p}{2}}=0
\ee
\be\label{E58}
d_1^2(d_2+d_p)\pm c.p. = A(d_1\pm c.p.)~,~d_1^2(d_3+d_{p-1})\pm c.p. = A(d_1\pm c.p.)
\ee
\be\label{E59}
d_1^2(d_{\frac{p}{2}}+d_{\frac{p}{2}+2})\pm c.p. = A(d_1\pm c.p.)~,
d_1^2 d_{\frac{p}{2}+1}\pm c.p. = \sqrt{1-m} (d_1 \pm c.p.)
\ee
\be\label{E60}
d_1d_2d_3 \pm c.p. = A(d_1\pm c.p.)~,~ d_1 d_j d_k \pm c.p. = A (d_1\pm c.p.)
\ee
\sss
\noindent{\bf p = Odd Integer:}

For indices $1<{j_1}<{j_2} \le p\,$:
\be\label{E61}
d_1d_{j_1}d_{j_2}+c.p.=A(d_1+c.p.)~,~{\tilde c}_1{\tilde c}_{j_1}{\tilde c}_{j_2}+c.p.=A({\tilde c}_1+c.p.)~,~{\tilde s}_1{\tilde s}_{j_1}{\tilde s}_{j_2}+c.p.=A({\tilde s}_1+c.p.)
\ee
\be\label{E62}
{\tilde c}_1({\tilde s}_2{\tilde d}_p+{\tilde s}_p{\tilde d}_2)+c.p. =0~~,\cdots,~~
{\tilde c}_1({\tilde s}_{\frac{p+1}{2}}{\tilde d}_{\frac{p+3}{2}}+{\tilde s}_{\frac{p+3}{2}}{\tilde d}_{\frac{p+1}{2}})+c.p. =0
\ee
\be\label{E63}
{\tilde c}_1{\tilde d}_2{\tilde d}_p+c.p.=A({\tilde c}_1+c.p.)~,\cdots,~~{\tilde c}_1{\tilde d}_{\frac{p+1}{2}}{\tilde d}_{\frac{p+3}{2}}+c.p.=A({\tilde c}_1+c.p.)
\ee
\be\label{E64}
{\tilde d}_1({\tilde d}_2{\tilde c}_2+{\tilde d}_p{\tilde c}_p)+c.p.=A({\tilde c}_1+c.p.)~,\cdots,~~{\tilde d}_1({\tilde d}_{\frac{p+1}{2}}{\tilde c}_{\frac{p+1}{2}}+{\tilde d}_{\frac{p+3}{2}}{\tilde c}_{\frac{p+3}{2}})+c.p.=A({\tilde c}_1+c.p.)
\ee
\be\label{E65}
d_1^2(d_2+d_p)+c.p. = A(d_1+c.p.)~,\cdots,~~d_1^2(d_{\frac{p+1}{2}}+d_{\frac{p+3}{2}})+c.p. = A(d_1+c.p.)
\ee
\be\label{E66}
c_1s_1(d_2+d_p)+c.p. = 0~,\cdots,~~c_1s_1(d_{\frac{p+1}{2}}+d_{\frac{p+3}{2}})+c.p. = 0
\ee
Note that additional identities can be obtained by changing the pair $({\tilde c},{\tilde d})$ in eq. (\ref{E63}) or eq. (\ref{E64})  into any of the pairs $({\tilde c},{\tilde s}),({\tilde s},{\tilde d}),({\tilde s},{\tilde c}),({\tilde d},{\tilde c}),({\tilde d},{\tilde s})$. Likewise, additional identities can be obtained by changing $d$ to ${\tilde c}$ or ${\tilde s}$ in eq. (\ref{E65}) and by changing $(c,s,d)$ to $({\tilde c},{\tilde d},{\tilde s})$ or $({\tilde d},{\tilde s},{\tilde c})$ in eq. (\ref{E66}).

\newpage
\noindent{\bf Table 3: Some identities of rank 4 and above.} The symbols $A,B,C$ in eqs. (\ref{E76}) through (\ref{E79}) are used generically to denote constants independent of $x$; the constants  
are in general all different. 

\sss
\noindent{\bf r = 4, p = 2:}
\be\label{E67}
d_1^3 d_2 \pm d_2^3 d_1 = \sqrt{1-m}\,(d_1^2 \pm d_2^2)~~,~~d_1^2 d_2^2=1-m
\ee
\be\label{E68}
m c_1 s_1 c_2 s_2 = \sqrt{1-m}[1-s_1^2 -s_2^2]~~,~~ c_1 d_1 c_2 d_2 = -(1-m)s_1 s_2~~
,~~s_1 d_1 s_2 d_2 = -c_1 c_2
\ee
\sss
\noindent{\bf r = 4, p = 3:}  $~~~[q \equiv \dn(2K(m)/3,m)]$
\be\label{E69}
s_1c_1d_2d_3+c.p. = \frac{(q^2+m-1)}{1-q^2} (s_1c_1+c.p.)
\ee
\be\label{E70}
d_1^3 (d_2+d_3)+c.p. = \frac{2mq}{1-q^2}(d_1^2+c.p.)-2(1-m)
\ee
\be\label{E71}
{\tilde s}_1 {\tilde d_1} {\tilde c}_2 {\tilde c}_3 + c.p. = 
\frac{q^2}{1-q^2} ({\tilde s}_1 {\tilde d}_1 + c.p.)
\ee
\be\label{E72}
{\tilde c}_1 {\tilde d_1} {\tilde s}_2 {\tilde s}_3 + c.p. = 
\frac{-1}{1-q^2} ({\tilde c}_1 {\tilde d}_1 + c.p.)
\ee
\be\label{E73}
m^2c_1s_1c_2s_2+c.p. = \frac{2mq}{1-q^2}(d_1^2+c.p.) +[m -(2-m)(1+q)^2] 
\ee
\sss
\noindent{\bf r = 5, p = 3:}
\be\label{E74}
d_1^3(s_2c_2+s_3c_3)+c.p.=\frac{-2mq}{1-q^2}(s_1 c_1 d_1+c.p.)
\ee
\be\label{E75}
m{\tilde s}_1^4({\tilde s}_2+{\tilde s}_3)+c.p. = 
\frac{2(q^2+m-1)}{1-q}({\tilde s}_1^3+c.p.)+\frac{2(q^3+q^2+mq-q+2m-1)}{(1-q^2)^2}({\tilde s}_1+c.p.)
\ee
\sss
\noindent{\bf r = 6, p = 6:}
\be\label{E76}
d_1^3(d_2^2d_3+d_6^2d_5)+c.p.=A(d_1^4+c.p.)+B(d_1^2+c.p.)+C~
\ee
\sss
\noindent{\bf r = 8, p = 6:}
\be\label{E77}
c_1d_1c_2d_2s_3s_4s_5s_6+c.p. = A(s_1s_2s_3s_4s_5s_6)
\ee
\sss
\noindent{\bf r, p:}
\be\label{E78}
m^ps_1^2 s_2^2 \cdots s_p^2 = A(s_1^2+c.p.)+B~~~(r=2p~,~p={\rm even})
\ee
For indices $1<{j_1}<{j_2}<\cdots<{j_{r-1}} \le p\,$:
$$
d_1 d_{j_1}d_{j_2} \cdots d_{j_{r-1}}+c.p. = A~~~(r={\rm even}~,~p)
$$
$$
d_1 d_{j_1}d_{j_2} \cdots d_{j_{r-1}}+c.p. = B(d_1+c.p.)~~~(r={\rm odd}~,~p)
$$
$$
{\tilde s}_1 {\tilde s}_{j_1}{\tilde s}_{j_2} \cdots {\tilde s}_{j_{r-1}}+c.p. = A~~,~~{\tilde c}_1 {\tilde c}_{j_1}{\tilde c}_{j_2} \cdots {\tilde c}_{j_{r-1}}+c.p. = A~~~(r={\rm even}~,~p={\rm odd})
$$
$$
{\tilde s}_1 {\tilde s}_{j_1}{\tilde s}_{j_2} \cdots {\tilde s}_{j_{r-1}}+c.p. = B({\tilde s}_1+c.p.)~~,~~{\tilde c}_1 {\tilde c}_{j_1}{\tilde c}_{j_2} \cdots {\tilde c}_{j_{r-1}}+c.p. = B({\tilde c}_1+c.p.)~~(r={\rm odd}~,~p={\rm odd})
$$
\be\label{E79}
d_1^{r-1}(d_2 + d_p) + c.p. = A(d_1^{r-2}+c.p.)+B(d_1^{r-4}+c.p.)+...~.
\ee
Eq. (\ref{E79}) ends with a constant if $r$ is odd and with a term  proportional to ($d_1+c.p.$), 
if $r$ is even.

\end{document}